\documentclass[pdftex]{article}
\usepackage{spconf,amsmath,amssymb,graphicx}
\usepackage{bmpsize}
\usepackage{algpseudocode,algorithm}
\usepackage{cite}
\usepackage{color}
\usepackage{comment}

\algnewcommand\algorithmicinput{\textbf{Input:}}
\algnewcommand\Input{\item[\algorithmicinput]}
\algnewcommand\algorithmicoutput{\textbf{Output:}}
\algnewcommand\Output{\item[\algorithmicoutput]}

\title{latexdiff Example - Revised version} 
\author{F Tilmann}
\RequirePackage[normalem]{ulem} 
\RequirePackage{color} 
\providecommand{\DIFadd}[1]{{\color{blue}\uline{#1}}} 
\providecommand{\DIFdel}[1]{{\color{red}\sout{#1}}} 
\providecommand{\DIFaddend}{} 
\providecommand{\DIFdelend}{} 

\title{CONVOLUTIONAL-SPARSE-CODED DYNAMIC MODE DECOMPOSITION \\ AND ITS APPLICATION TO RIVER STATE ESTIMATION}
\name{
  Y.~Kaneko$^{\dag}$,
  S.~Muramatsu$^{\ast}$,
  H.~Yasuda$^{\ddag}$,
  K.~Hayasaka$^{\dag\dag}$,
  Y.~Otake$^{\ast}$,
  S.~Ono$^{\S}$ and
  M.~Yukawa$^{\P}$
\thanks{This work was supported by a U-go Grant of Niigata University.}}%
\address{
\dag Graduate School of Sci. \& Tech., Niigata Univ., Japan,\\
\ddag Research Inst. for Natural Hazard \& Disaster Recovery, Niigata Univ., Japan,\\
\dag\dag Faculty of Sci., Niigata Univ., Japan, \
\S Inst. of Innovative Research, Tokyo Institute of Tech., Japan, \\
$\ast$ Faculty of Eng., Niigata Univ., Japan, \
\P Faculty of Sci. \& Tech., Keio Univ., Japan
}

\begin{document}
\ninept
\maketitle

\begin{abstract}
This work proposes convolutional-sparse-coded dynamic mode decomposition (CSC-DMD) by unifying extended dynamic mode decomposition (EDMD) and convolutional sparse coding.
EDMD is a data-driven method of analysis used to describe a nonlinear dynamical system with a linear time-evolution equation. 
 Compared with existing EDMD methods, CSC-DMD has the advantage of reflecting the spatial structure of a target.
 As an example, the proposed method is applied to river bed shape estimation from the water surface observation. This estimation problem is reduced to sparsity-aware signal restoration with a hard constraint given by the CSC-DMD prediction, where the  algorithm is derived by the primal-dual splitting method. A time series set of water surface and bed shape measured through an experimental river setup is used to train and test the system.
 From the result, the efficacy of the proposed method is verified.

\end{abstract}

\section{Introduction}
\label{sec:intro}

Recent developments in sensing, networking and computer technology have increased the appeal of modeling complex dynamical phenomena.
In particular, modeling of high-dimensional and nonlinear dynamical systems is required in many fields in engineering, physics and biology. Even when analytical modeling is difficult, we can obtain time evolution equation of a dynamical system by using a data-driven approach with training data if some low-dimensional spatiotemporal coherence structure dominates the phenomenon.
Extended dynamic mode decomposition (EDMD) is one such data-driven approaches and applied to various fields such as fluid dynamics analysis\cite{dmdnu,ondmd,dmd,edmd}. The technique can be viewed as a finite dimensional approximation of the Koopman mode decomposition (KMD), which expresses a nonlinear dynamical system by a linear time evolution in infinite dimensionality\cite{kmd}.
%

Many residential areas worldwide have suffered from river flooding\cite{flood}.
From this background, we try to 
measure river state in high resolution with low cost\cite{weihang}.
The technology for achieving this requires a time-domain formula to allow the detection of abnormal conditions and to control the flow path. Since the mathematical model representing the river dynamics is high dimensional and nonlinear, we adopt EDMD for its modeling.
Existing EDMD techniques, however, require an analysis dictionary, which implies significant challenges.
%
%
Williams et al. proposed EDMD with kernel\cite{edmd} and Yeung et al. proposed EDMD with neural networks\cite{nndmd}.
However, these EDMD methods are not necessarily intuitive to incorporate spatial structures of the target phenomena due to their analysis point of view.
Such a feature may prevent EDMD from a wide range of applications.

We therefore propose an approach to EDMD.
To simplify the description of a spatial structure, we adopt convolutional sparse coding (CSC) to constitute a synthesis dictionary\cite{csc2017,csr,cdl,mlconv}.
Since spatial local variations of river surface and bottom are closely related to each other, it is expected to represent them sparsely with CSC. In addition, there is a merit that it is easy to reflect the domain knowledge to the dictionary structure.
The time-evolution formula is derived by combining CSC and EDMD (CSC-DMD).
The dictionary is trained by a data set acquired through a river experimental setup, which we refer to as a stream tomography (ST) system\cite{hoshino}.
Assuming that the shape of the water can be measured,
we formulate the river bed estimation task as a minimization problem of the data fidelity penalized by the $\ell_1$ norm under a hard constraint specified by the CSC-DMD formula
and give a solver through the primal-dual splitting (PDS) method\cite{Condat2013,hcopds,pds}.
To verify the CSC-DMD outcome, we evaluate the estimation performance by using test data acquired by ST and compared with the results with normal DMD.

\section{Review of EDMD}
\label{sec:edmd}

Let us review EDMD as preparation.
%
%
This method approximates KMD with a finite dimensional expression\cite{kmd,edmd} and
decomposes high-dimensional dynamics into multiple spatial modes with temporal variations. It enables us to understand spatial structure and temporal evolution of a given set of time-series data.

Let  $\mathbf{F}$ : $\mathcal{M} \to \mathcal{M}$ be a nonlinear map that relates a pair of snap shots in finite dimensional discrete time series data $\{ \mathbf{x}_k\in\mathcal{M}\subseteq\mathbb{R}^M \}^{N-1}_{k=0}$ such as $\mathbf{x}_k = \mathbf{F} \bigl( \mathbf{x}_{k-1} \bigl)$, where $\mathcal{M}$ denotes a state space.
If there is a function space\footnote{In \cite{edmd}, the function space $\mathcal{F}$ is referred to as observable space.} $\mathcal{F} \ni \psi\colon \mathcal{M} \to \mathbb{R}$, and  exists an operator $\mathcal{K}$ : $\mathcal{F} \to \mathcal{F}$ such that
    $\mathcal{K} \psi = \psi \circ  \mathbf{F}$,
then, $\mathcal{K}$ is refered to as a the Koopman operator of $\mathcal{F}$.

In EDMD, a nonlinear map $\mathbf{\Psi} \colon \mathcal{M} \to \tilde{\mathcal{F}}$ is introduced as
\begin{equation}
    \mathbf{y}_k = \mathbf{\Psi} \bigl( \mathbf{x}_k \bigl)
\end{equation}
for $\{ \mathbf{x}_k\}^{N-1}_{k=0}$, where
$\mathcal{\tilde{F}}\subseteq\mathbb{R}^L$ and $M \leq L$.
The time evolution of $\mathbf{y}_k$ is approximated by a matrix $\mathbf{K}\colon \tilde{\mathcal{F}} \to \tilde{\mathcal{F}}$ as
\begin{equation}
    \mathbf{y}_k \approx \mathbf{K}\mathbf{\Psi} \bigl( \mathbf{x}_{k-1} \bigl) = \mathbf{K}\mathbf{y}_{k-1},
\end{equation}
where we refer to $\tilde{\mathcal{F}}$ as feature space.
$\tilde{\mathcal{F}}$ can be interpreted as a space of coefficient vectors which contain a finite number of scalers
multipled to eigen vectors of subspace of $\mathcal{F}$ to approximate $\psi\in\mathcal{F}$ with linear combination.

To derive the matrix $\mathbf{K}$, data matrix
\begin{equation}
    \mathbf{Y}_{k\colon \ell} \triangleq \begin{pmatrix}
    \mathbf{y}_k & \mathbf{y}_{k+1} & \cdots & \mathbf{y}_{\ell}
  \end{pmatrix} \in \mathbb{R}^{L\times (\ell-k+1)}
  \label{eq:Ykl}
\end{equation}
is defined by the coefficient vectors $\{ \mathbf{y}_k\in\tilde{\mathcal{F}} \}^{N-1}_{k=0}$, where
$k \leq \ell \leq N-1$.
Then, $\mathbf{K}$ is approximately obtained by
\begin{equation}
\mathbf{K} = \mathbf{Y}_{1\colon N-1}\mathbf{Y}_{0\colon N-2}^{+}
\end{equation}
as a solution of
%
$\mathbf{K} \in \arg\min_{\mathbf{A}}\|\mathbf{Y}_{1\colon N-1}-\mathbf{A}\mathbf{Y}_{0\colon N-2}\|_{\mathrm{F}}^2$,
%
where $\|\cdot\|_{\mathrm{F}}$ represents the Frobenius norm and  superscript ${+}$ represents the pseudo-inverse of Moore-Penrose.

Matrix $\mathbf{K}$ is decomposed by the following procedure \cite{dmd,edmd}.
First, the singular value decompostion (SVD) is applied to $\mathbf{Y}_{0: N-2}$ to obtain $\mathbf{U}_r \in \mathbb{C}^{L \times r}$,
$\mathbf{\Sigma}_r \in \mathbb{C}^{r \times r}$,
$\mathbf{V}_r \in \mathbb{C}^{(N-1) \times r}$, where $r\leq\mathrm{rank}(\mathbf{Y}_{0\colon N-2})$.
Let here $\mathbf{\check{K}} = \mathbf{U}^{\ast}_r \mathbf{Y}_{1\colon N-1} \mathbf{V}_r \mathbf{\Sigma}^{-1}_r$, which is a projection of $\mathbf{K}$ to $\mathbf{U}_r$. Then, we obtain
$\mathbf{W}\in\mathbb{C}^{r\times r}$
such that $\mathbf{\check{K}}\mathbf{W} = \mathbf{W}\mathbf{\Lambda}$, where $\mathbf{\Lambda}\in\mathbb{C}^{r\times r}$ represents a diagonal matrix, superscript ${\ast}$ means Hermitian transposition.
Since $\mathbf{K}$ and $\mathbf{\check{K}}$ are similar, $\mathbf{\Lambda}$ consists of eigenvalues of $\mathbf{K}$.
The dynamic mode $\mathbf{\Phi}$ of $\mathbf{K}$ is obtained as
\begin{equation}
    \mathbf{\Phi} = \mathbf{Y}_{1\colon N-1} \mathbf{V}_r \mathbf{\Sigma}^{-1}_r \mathbf{W}.
\end{equation}
As a result, an approximation of time evolution equation is given by
\begin{equation}
    \mathbf{y}_k \approx \mathbf{\Phi} \mathbf{\Lambda}^{k} \mathbf{b}_0,
    \label{eq:yk2}
\end{equation}
where $\mathbf{b}_0=\mathbf{\Phi}^{+}\mathbf{y}_0$.
The temporal evolution formula of the original time series data $\{\mathbf{x}_k\}_{k=0}^{N-1}$ is obtained by
\begin{equation}
  {\mathbf{x}}_k\approx
  \mathbf{V}\mathbf{\Lambda}^k\mathbf{b}_0,
  \label{eq:xk1}
\end{equation}
where $\mathbf{V}=\tilde{\mathbf{D}}\mathbf{\Phi}$,
%
 $\tilde{\mathbf{D}}
  = \arg\min_{\mathbf{D}}\|\mathbf{X}_{0\colon N-1}-\mathbf{D}\mathbf{Y}_{0\colon N-1}\|_\mathrm{F}^2$ and
$\mathbf{X}_{k\colon \ell} \triangleq \begin{pmatrix}
    \mathbf{x}_k & \mathbf{x}_{k+1} & \cdots & \mathbf{x}_{\ell}
  \end{pmatrix} \in \mathbb{R}^{M\times (\ell-k+1)}$.
%
Note that $\tilde{\mathbf{D}}\colon \tilde{\mathcal{F}}\rightarrow \mathcal{M}$ has a role of a synthesis dictionary.

Let $t$ be continuous time and $\bigtriangleup t$ be the sampling interval. Then, the continuous version of the time evolution equation is given by
\begin{equation}
  \mathbf{x}(t)=\mathbf{V} e^{\mathbf{\Omega}t} \mathbf{b}_0,
  \label{eq:xtime}
\end{equation}
where $\mathbf{\Omega} = {\ln{\mathbf{\Lambda}}}/{\Delta t}$.
$\mathbf{\Omega}$ and $\mathbf{\Lambda}$ are diagonal matrices, and $\mathrm{ln}(\cdot)$ takes a logarithm for each diagonal element.

For existing EDMD techniques, an appropriate selection of map $\mathbf{\Psi}$ is crucial\cite{edmd}.
Although methods such as using kernels\cite{edmd} and neural networks\cite{nndmd} are proposed, how to incorporate structure suitable for spatial relations into the mapping is nontrivial.
In the above approach, there remains a problem how to incorporate spatial structures into map $\mathbf{\Psi}$.

\section{Convolutional-Sparse-Coded DMD}\label{sec:cscdmd}

\begin{figure}[tb]
\centering
\includegraphics[width=.7\linewidth]{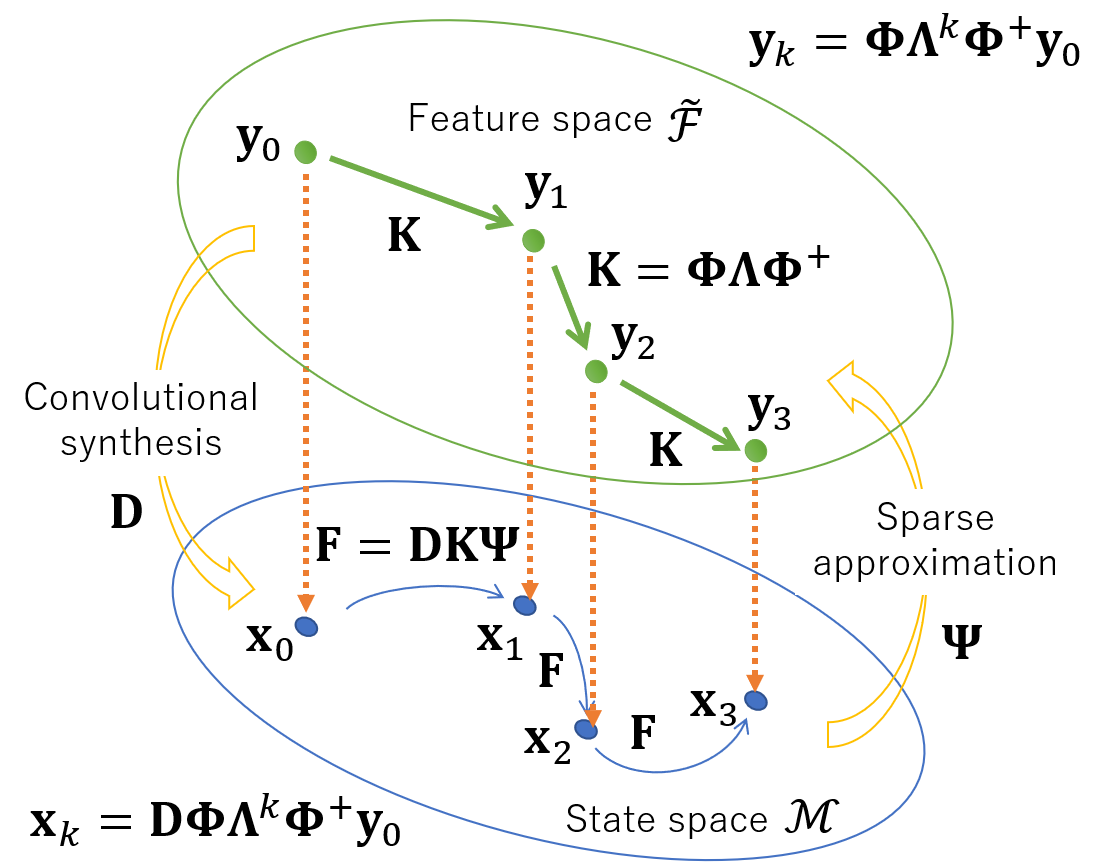}
\caption{Basic concept of proposed CSC-DMD.}\label{fig:csdmd}
\end{figure}

In this section, we propose an EDMD with a synthesis-dictionary-oriented approach to make appropriate model setting simpler. First, a synthesis dictionary $\mathbf{D}$ is obtained, which satisfies $\mathbf{X}_{0\colon N-1}\approx \mathbf{D}\mathbf{Y}_{0\colon N-1}$, by using CSC instead of giving an analysis map $\mathbf{\Psi}$ as in Sect.~\ref{sec:edmd}\cite{csc2017,csr,cdl,mlconv}. Fig.~\ref{fig:csdmd} summarizes the basic idea. The problem setting of CSC can be formulated as
\begin{gather}
  \{\hat{\mathbf{D}},\{\hat{\mathbf{y}}_k\}\}=\arg\min_{\mathbf{D},\{\mathbf{y}_k\}} J(\mathbf{D},\{\mathbf{y}_k\}\ \vert\ \{\mathbf{x}_k\}),\label{eq:csc}
\\
J(\mathbf{D},\{\mathbf{y}_k\} \vert \{\mathbf{x}_k\}) \triangleq
  \sum_{k=0}^{N-1}\left(\frac{1}{2}\|\mathbf{x}_k-\mathbf{Dy}_k\|_2^2 + \lambda\rho(\mathbf{y}_k)\right),
\end{gather}
where $J$ is a cost function,  $\rho$ : $\tilde{\mathcal{F}} \to [0,\infty$) is a regularization term and $\lambda \geq 0$ is a regularization parameter.
Eq.~\eqref{eq:csc} is a simultaneous optimization problem of dictionary $\mathbf{D}$ and sparse coefficient vectors $\{ \mathbf{y}_k \in \mathcal{\tilde{F}} \}^{N-1}_{k=0}$.
As typical dictionary learning, we alternately apply a sparse approximation and dictionary update step until some stopping criterion is satisfied. These steps are summarized as follows:
\begin{description}
\item [Sparse approximation step]
Fix a synthesis dictionary $\mathbf{\Hat{D}}$ and obtain the sparse coefficients
\begin{multline}
\hat{\mathbf{y}}_k=\arg\min_{\mathbf{y}\in\mathbb{R}^L}\frac{1}{2}\|\mathbf{x}_k-\hat{\mathbf{D}}\mathbf{y}\|_2^2 + \lambda\rho(\mathbf{y}),
\\
k =0,1,2,\cdots,N-1.\label{eq:sprsaprxstep}
\end{multline}
%
%
\item [Dictionary update step] Fix the set of sparse coefficient vectors
$\{\hat{\mathbf{y}_k} \}^{N-1}_{k=0}$ and update the synthesis dictionary as
\begin{equation}
\hat{\mathbf{D}}=\arg\min_{\mathbf{D}\in\mathbb{R}^{M\times L}}\frac{1}{2}\sum_{k=0}^{N-1}\|\mathbf{x}_k-\mathbf{D}\hat{\mathbf{y}}_k\|_2^2 .
\label{eq:dicupd}
\end{equation}
\end{description}

We adopt a multidimensional filter bank with design parameters as a convolutional synthesis dictionary $\mathbf{D}$. In this case, \eqref{eq:dicupd} can be rewritten as
 $\hat{\mathbf{D}}=\mathbf{D}_{\hat{\mathbf{\Theta}}}$,
$\hat{\mathbf{\Theta}}=\arg\min_{\mathbf{\Theta}}\frac{1}{2}\sum_{k=0}^{N-1}\|\mathbf{x}_k-\mathbf{D}_{\mathbf{\Theta}}\hat{\mathbf{y}}_k\|_2^2$,
%
where  $\mathbf{\Theta}$ denotes a set of the design parameters.
For example, a non-separable oversampled lapped transform (NSOLT) can be used \cite{nsolt} and yields a convolutional dictionary that simultaneously satisfies the overlapping, redundant, symmetric, Parseval tightness and no-DC leakage property.

Once we obtain a synthesis dictionary $\hat{\mathbf{D}}$, we can define a nonlinear map $\mathbf{\Psi}$ by the following sparse approximation:
\begin{equation}
\mathbf{\Psi}(\mathbf{x})=\arg\min_{\mathbf{y}\in\mathbb{R}^L}\frac{1}{2}\|\mathbf{x}-\hat{\mathbf{D}}\mathbf{y}\|_2^2+\lambda\rho(\mathbf{y}),\label{eq:cscdmdpsi}
\end{equation}
and can follow the EDMD steps. When $\rho(\cdot)=\|\cdot\|_1$, the iterative soft-thresholding algorithm (ISTA) can be used to solve \eqref{eq:cscdmdpsi}\cite{ista}.

\section{Application to river state estimation}

\begin{figure}
    \centering
    \includegraphics[width=\linewidth]{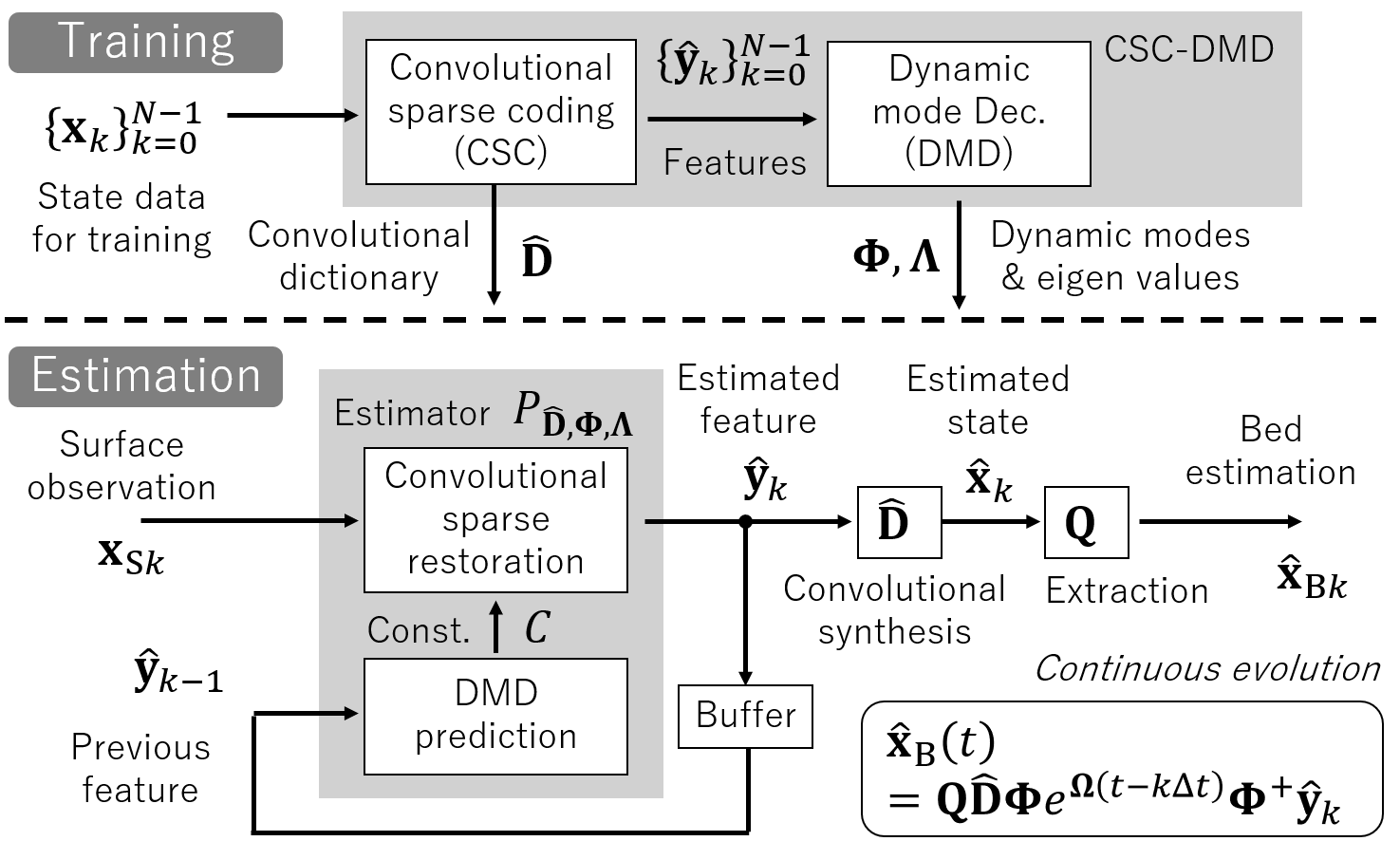}
    \caption{Diagram of the proposed river bed estimation system}
    \label{fig:estimator}
\end{figure}

In this section, we further propose a method to estimate invisible river bed shape to follow temporal variation of the flow path on the assumption that the river surface is visible. We formulate the problem as signal restoration with a hard constraint. Let us apply the proposed CSC-DMD prediction to constrain the restoration result and develop an algorithm for solving the problem through the PDS method\cite{Condat2013,hcopds,pds}.
Fig.~\ref{fig:estimator} shows a schematic diagram of the river bed estimation system.

\subsection{Problem setting of river bed estimation}

Let us consider the problem of estimating an invisible bottom shape $\mathbf{x}_{\mathrm{B}k} \in \mathbb{R}^{m_{\mathrm{B}}}$ from visible water surface shape $\mathbf{x}_{\mathrm{S}k} \in \mathbb{R}^{m_{\mathrm{S}}}$, where
 $\mathbf{x}_{\mathrm{S}k}$ and $\mathbf{x}_{\mathrm{B}k}$ are
assumed to exist at a time instance $t = k \triangle t$.
Here, we can define a state space vector $\mathbf{x}_k$ as
\begin{equation}
  \mathbf{x}_k \triangleq \begin{pmatrix}
      \mathbf{x}_{\mathrm{S}k} \\
      \mathbf{x}_{\mathrm{B}k}
  \end{pmatrix}\in\mathbb{R}^{M},
\end{equation}
where $M = m_{\mathrm{S}} + m_{\mathrm{B}}$.
Provided that time series data $\{\mathbf{x}_k\}^{N-1}_{k=0}$ is given as training data, we can obtain a time evolution equation \eqref{eq:xtime} through CSC-DMD.
In the followings, we discuss to construct an estimator $\mathcal{\tilde{P}_{\mathbf{\Hat{D},\Phi,\Lambda}}} \colon$ $\mathbb{R}^{m_\mathrm{B}} \times \mathbb{R}^L \to \mathbb{R}^L$, which realizes the bottom shape estimation as
\begin{align}
\hat{\mathbf{x}}_{\mathrm{B}k}&=\mathbf{Q}\hat{\mathbf{D}}\hat{\mathbf{y}}_k\\
  \hat{\mathbf{y}}_k&=\mathcal{P}_{\hat{\mathbf{D}},\mathbf{\Phi},
  \mathbf{\Lambda}}(\mathbf{x}_{\mathrm{S}k},\hat{\mathbf{y}}_{k-1}),
\end{align}
where $\hat{\mathbf{x}}_{\mathrm{B}k}$ and $\hat{\mathbf{y}}_{k}$ are estimates of river bed shape and feature vector at  time $k$, respectively, and $\mathbf{Q}$ is the matrix that extracts $\mathbf{x}_{\mathrm{B}k}$ from vector $\mathbf{x}_{k}$ defined by
$\mathbf{Q}\triangleq\begin{pmatrix}
\mathbf{O} & \mathbf{I}
\end{pmatrix}$, where $\mathbf{O}$ and $\mathbf{I}$ are the zero and identity matrices, respectively.
By using $\hat{\mathbf{b}}_{k} = \mathbf{\Phi}^{+}\hat{\mathbf{y}}_k$, we can update the time evolution equation of river bed as
\begin{equation}
\hat{\mathbf{x}}_{\mathrm{B}\ell} = \mathbf{Q}\mathbf{V}\mathbf{\Lambda}^{\ell-k}\hat{\mathbf{b}}_k,
\end{equation}
for $\ell\geq k$, that is,
\begin{equation}
\hat{\mathbf{x}}_\mathrm{B}(t) = \mathbf{Q}\mathbf{V}{e}^{\mathbf{\Omega}(t-k\Delta t)}
 \hat{\mathbf{b}}_k,
 \label{eq:contevol}
\end{equation}
for $t\geq k\Delta t$.

\subsection{Sparse signal restoration with CSC-DMD prediction}

\begin{algorithm}[t]
{\small
\caption{CSC-DMD predicted sparse signal restoration}\label{alg:hcist}
\begin{algorithmic}[1]
\Input $\mathbf{x}_{\mathrm{S}}$, $\mathbf{y}^{(0)}$, $\mathbf{q}^{(0)}$
\Output $\hat{\mathbf{x}}_{\mathrm{B}}$, $\hat{\mathbf{y}}$
\State{$n \gets 0$}
\State{$\mathbf{x}^{(0)} =\hat{\mathbf{D}}\mathbf{y}^{(0)}$}
\While{A stopping criterion is not satisfied}
 \State{$\mathbf{g} \gets \hat{\mathbf{D}}^\intercal
 (\mathbf{P}^\intercal(\mathbf{Px}^{(n)}-\mathbf{x}_{\mathrm{S}})+\mathbf{q}^{(n)})$}
\State{$\mathbf{y}^{(n+1)} = \mathfrak{G}_{\rho}(\mathbf{y}^{(n)}-\gamma_1\mathbf{g},\sqrt{\gamma_1\lambda})$}
\State{$\mathbf{x}^{(n+1)}=\hat{\mathbf{D}}\mathbf{y}^{(n+1)}$}
\State{$\mathbf{u} \gets 2\mathbf{x}^{(n+1)} - \mathbf{x}^{(n)}$}
\State{$\mathbf{z} \gets \mathbf{q}^{(n)}+\gamma_2\mathbf{u}$}
\State{$\mathbf{q}^{(n+1)} = \mathbf{z} -\gamma_2P_{C}({\gamma_2}^{-1}\mathbf{z})$}
\State{$n\gets n+1$}
\EndWhile
\State{$\hat{\mathbf{x}}_{\mathrm{B}}=\mathbf{Q}\mathbf{x}^{(n)}$}
\State{$\hat{\mathbf{y}}=\mathbf{y}^{(n)}$}
\end{algorithmic}
}
\end{algorithm}

For the river bed estimation, we define a sparsity-aware signal restoration model with a norm ball constraint determined by the CSC-DMD prediction in \eqref{eq:yk2}. The constraint takes the distance between CSC-DMD prediction and current restoration into account.
Considering current observation and past estimation, we give the estimator $\mathcal{P}_{\hat{\mathbf{D}},\mathbf{\Phi},
  \mathbf{\Lambda}}(\mathbf{x}_{\mathrm{S}k},\hat{\mathbf{y}}_{k-1})$ as
\begin{multline}
\hat{\mathbf{y}}_k=\arg\min_{\mathbf{y}\in\mathbb{R}^L}\frac{1}{2}\|\mathbf{x}_{\mathrm{S}k}-\mathbf{P}\hat{\mathbf{D}}\mathbf{y}\|_2^2+\lambda\rho(\mathbf{y}),\\
  \mathrm{s.t.}\ \|\hat{\mathbf{D}}(\mathbf{\Phi\Lambda}\mathbf{\Phi}^{+}\hat{\mathbf{y}}_{k-1}-\mathbf{y})\|_2^2\leq\varepsilon^2,
  \label{eq:problem}
\end{multline}
where $\mathbf{P}\triangleq\begin{pmatrix}
\mathbf{I} & \mathbf{O}
\end{pmatrix}$
and $\varepsilon\in[0,\infty)$ determines the tolerance of the prediction error.

\subsection{River bed estimation algorithm}

In order to solve the problem in \eqref{eq:problem}, we adopt the PDS method\cite{Condat2013,pds,agppds,hcopds}.  {\bf{Algorithm 1}} shows the derived algorithm, where
$\mathfrak{G}_{\rho}(\cdot,\sigma)\colon \mathbb{R}^L\rightarrow\mathbb{R}^L$ is a Gauss denoiser with regularizer $\rho$\cite{pds}. $\gamma_1$ and $\gamma_2$ are step size parameters, which are set to hold $\gamma_1^{-1}-\gamma_2(\sigma_1(\mathbf{\Hat{D}}))^2 \geq \beta/2$, where $\sigma_1(\cdot)$ means the maximum singular value and $\beta$ is the Lipschitz constant of the gradient of the fidelity term, i.e., $\beta=(\sigma_1(\mathbf{P}\hat{\mathbf{D}}))^2$\cite{pds}.
Note that if the synthesis dictionary $\Hat{\mathbf{D}}$ satisfies the Parseval tightness $\mathbf{\Hat{D}\Hat{D}}^{\intercal} = \mathbf{I}$, then  $\sigma_1(\mathbf{\Hat{D}}) = 1$ and $\sigma_1(\mathbf{P}\hat{\mathbf{D}})=1$ hold.
Non-separable oversampled lapped transform can guarantee this property\cite{nsolt}.

The optimization problem in \eqref{eq:problem} is rewritten as
\begin{equation}
\hat{\mathbf{y}}=\arg
  \min_{\mathbf{y}\in\mathbb{R}^{L}}
  f_{\mathbf{x}_\mathrm{S}}(\mathbf{P}\hat{\mathbf{D}}\mathbf{y})+
\lambda\rho(\mathbf{y}) + \imath_C(\hat{\mathbf{D}}\mathbf{y}),
\end{equation}
where $f_{\mathbf{v}}:\mathbb{R}^{m_\mathrm{S}}\rightarrow [0,\infty)$ is a data fidelity term, $\rho\colon\mathbb{R}^L\rightarrow[0,\infty)$ is a regularization term, $\imath_C\colon \mathbb{R}^{M}\rightarrow \{0,\infty\}$ is an indicator function of a constraint set $C$, and they have the following forms:
%
 $f_{\mathbf{v}}(\mathbf{x}) = \frac{1}{2}\|\mathbf{v}-\mathbf{x}\|_2^2$,
 $\rho(\mathbf{y})= \|\mathbf{y}\|_1$ 
and $C=\{\mathbf{x}\in\mathbb{R}^M\ \vert\ \|\mathbf{c}-\mathbf{x}\|_2^2
  \leq\varepsilon^2\}$,
%
where $\mathbf{c}=\hat{\mathbf{D}}\mathbf{\Phi\Lambda\Phi}^{+}\hat{\mathbf{y}}_{k-1}\in\mathbb{R}^M$.

Since $\rho(\cdot)=\|\cdot\|_1$, the Gaussian denoiser $\mathfrak{G}_{\rho}(\cdot,\sigma)$ is reduced to the soft thresholding
\begin{equation}
  \mathfrak{G}_{\|\cdot\|_1}(\mathbf{x},\sigma)=\mathrm{sgn}(\mathbf{x})\odot
  \max\left(\mathrm{abs}(\mathbf{x})-\sigma^2\mathbf{1},\mathbf{0}\right),  \label{eq:softthresh}
\end{equation}
where $\mathrm{sgn}(\cdot)$ and $\mathrm{abs}(\cdot)$ return the sign and the absolute value for each element, respectively, $\max(\cdot,\cdot)$ returns the maximum value element-wisely,  $\odot$ denotes the Hadamard product,
$\mathbf{0}$ and $\mathbf{1}$ are vectors of all zeros and all ones, respectively.
The metric projection $P_C(\cdot)$ is given as
\begin{equation}
  P_C(\mathbf{x})=\left\{\begin{array}{ll}
\mathbf{x}, & \mathbf{x}\in C,\\
\mathbf{c}+
\frac{\varepsilon}{
\|\mathbf{x}-\mathbf{c}\|_2}(\mathbf{x}-\mathbf{c}), &\mathrm{otherwise}
\end{array}\right..
\end{equation}
\section{Performance evaluation} 
\begin{figure}[tb]
    \centering
    \includegraphics[width=.85\linewidth]{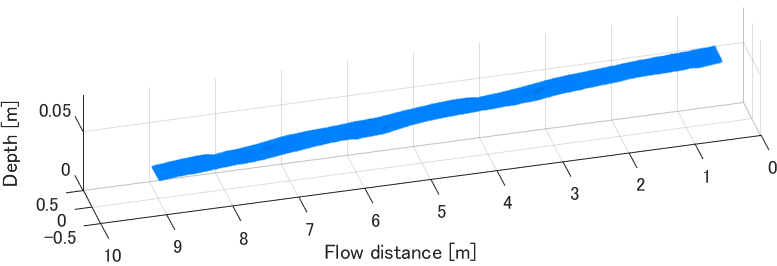}\\
    { (a)}\\
        \includegraphics[width=.85\linewidth]{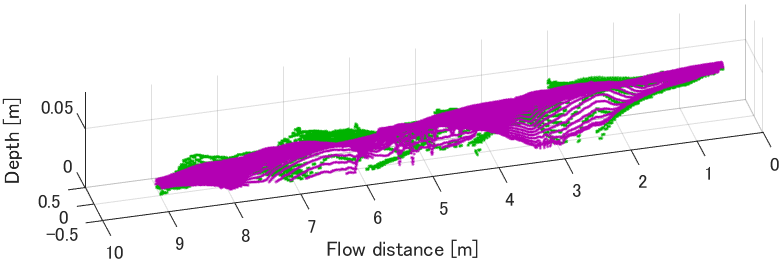}\\
        { (b)}\\
    \caption{
    River state measured by ST at 440 minutes after the start of water and sand inflow.  (a) on the water surface, (b) on the river bed, where the measurement is in magenta and the estimation is in green ($\varepsilon=0.02$). The number of points are 41$\times$851.
    }
    \label{fig:stdata}
\end{figure}

\begin{figure}[tb]
    \centering
    \includegraphics[width=1\linewidth]{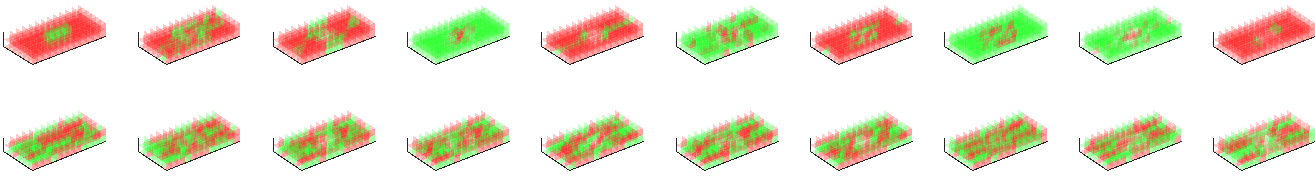}
    \caption{Convolutional kernels designed by Type-I NSOLT dictionary learning, where decimation factor  $[M_\mathrm{y}, M_\mathrm{x}, M_\mathrm{z}] = [2,4,2]$, polyphase order $[N_\mathrm {y}, N_\mathrm{x},N_\mathrm{z}]=[2,2,0]$, and the number of channels $P=20$. Every filter kernel has symmetry and is of size $6 \times 12 \times 2$. Parseval tightness is satisfied and there is no DC leakage. Green and red means positive and negative value, respectively.}
    \label{fig:nsoltdic}
\end{figure}

In this section, in order to show the significance of the proposed method, we evaluate the estimation performance.

\subsection{Experimental river setup}

Hoshino et al. developed a river experimental setup to simultaneously measure the shape of water surface and bottom sediment to understand the sand transportation mechanism in river\cite{hoshino}. The ST system scans flowing water and moving sand by using sheet laser to detect motion from the light reflections.
The experimental river model is of length 12m and of width 0.45m, and the water path gradient is $1/200$ and the flow rate is 2.0$\ell$/sec.
Fig.~\ref{fig:stdata}~(a) shows an example of the measured water surface,
and the corresponding bottom shape is shown in Fig.~\ref{fig:stdata}~(b) in magenta.

As the training data for CSC-DMD, the ST measurement data of water surface and bed shape from 100 to 250 minutes after the start of inflow are used, where the interval is set to 10 minutes.
In addition, the measurement data from 260 to 440 minutes are used as test data for evaluating the bottom shape estimation.

\begin{figure}[tb]
  \centering
  \begin{minipage}{.45\linewidth}
    \centering
    \includegraphics[width=.9\linewidth]{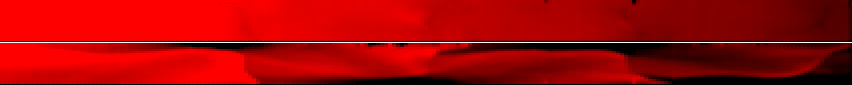}\\
        { (a)}\\
        \smallskip
    \includegraphics[width=.9\linewidth]{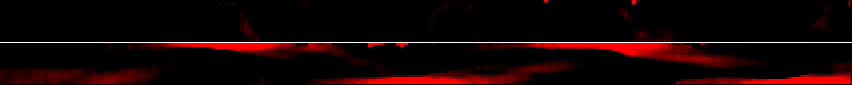}\\
     { (b)}\\
        \smallskip
            \includegraphics[width=.9\linewidth]{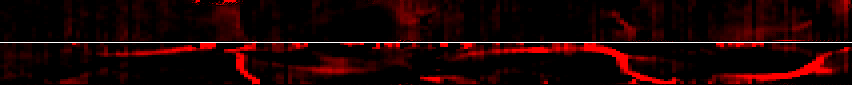}\\
        { (c)}\\
        \smallskip
    \includegraphics[width=.9\linewidth]{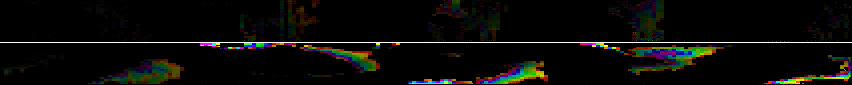}\\
     { (d)}\\
        \smallskip
    \includegraphics[width=.9\linewidth]{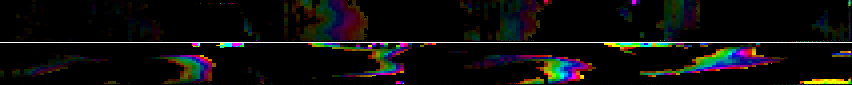}\\
        { (e)}
\end{minipage}
  \begin{minipage}{.45\linewidth}
    \centering
\includegraphics[width=\linewidth]{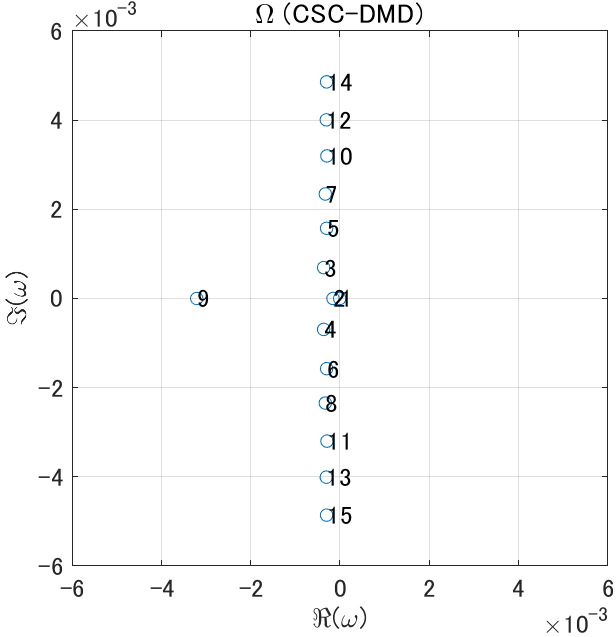}\\
 {\small (f)}
  \end{minipage}
  \caption{Examples of dynamic modes in $\mathbf{V}$ and placement of the eigenvalues of CSC-DMD.
  (a) constant mode 1, (b)(c) decay mode 2,9, (d)(e) vibration decay mode 3,4, where the upper and lower half potion correspond to the water surface and bed shape, respectively, and hue corresponds to the phase and red means real number. (f) eigenvalues in $\mathbf{\Omega}$. }
  \label{fig:modeetc}
\end{figure}

\begin{figure}
    \centering
    \includegraphics[width=.8\linewidth]{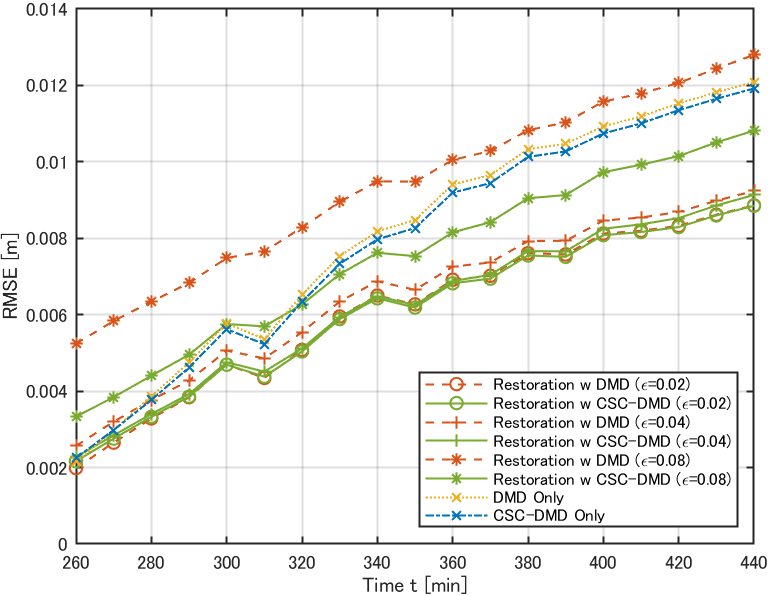}
    \caption{Experimental result of river bed estimation}
    \label{fig:bedest}
\end{figure}

\subsection{Result of CSC-DMD learning}

Fig.~\ref{fig:nsoltdic} shows convolutional kernels trained for ST data, and Fig.~\ref{fig:modeetc} shows some of dynamic modes and placement of eigenvalues in $\mathbf{\Omega}$ learned by CSC-DMD.
Type-I NSOLT is adopted to construct a convolutional dictionary $\mathbf{D}$\cite{nsolt}. 
The Parseval tightness and no-DC leakage properties are set to hold.
Every filter has symmetry and is of size $6\times 12\times 2$.   Atom termination is applied to the boundary\cite{furuya,boundly}.
The sparse approximation in \eqref{eq:sprsaprxstep} and \eqref{eq:cscdmdpsi} used ISTA\cite{ista}.
In the dictionary update step \eqref{eq:dicupd}, a stochastic gradient descent method with AdaGrad was adopted \cite{adagrad,nsolticip,fujii}, where 128 patches of size $32\times 128\times 2$ voxels were randomly extracted from the ST training data.

%
\vspace{-2mm}
\subsection{Experimental results of river bed estimation}

Let us evaluate the river bed estimation performance of the proposed method.
We compare the results of the CSC-DMD-predicted sparse signal restoration with those of the normal-DMD-predicted one.
The latter corresponds to the CSC-DMD with the identity dictionary $\mathbf{D}=\mathbf{I}$.
Fig.~\ref{fig:bedest} shows the root mean squared error (RMSE) results for the test data, where $\varepsilon \in \{0.02,0.04,0.08\}$, $\lambda=8\times 10^{-3}$, $\gamma_1 = 1\times 10^{-3}$ and  $\gamma_2 = 950$.
The CSC-DMD prediction (green) performs better than the normal DMD (red), and the proposed restoration method also performs better than the bare DMD prediction (yellow) and the bare CSC-DMD prediction (blue).

\section{Conclusions}

In this work, we proposed  applying CSC to EDMD.
In order to verify the significance of the proposed method, we tried to adopt the method for estimating a time series of invisible river bed shape from a sequence of visible river surface snapshots. The proposed method was evaluated for data acquired by a river experimental setup.
Demonstrably, the proposed method performs better than the DMD approach.
Future work will investigate the tolerance to state change, spatial transition and multi-scaleability.

%


\newpage

\end{document}